\documentclass[12pt]{article}
\newcommand{\be}{\begin{eqnarray}}\newcommand{\ee}{\end{eqnarray}}
\newcommand{\beo}{\begin{eqnarray*}}\newcommand{\eeo}{\end{eqnarray*}}
\newcommand{\ba}{\begin{array}}\newcommand{\ea}{\end{array}}
\newcommand{\no}{\nonumber}
\begin{document}
\baselineskip=18pt
\parskip=3pt
\begin{titlepage}
\title{Tachyon Condensation and Background\\Independent Superstring Field Theory}

\author{K. S. Viswanathan\footnote{kviswana@sfu.ca}\hspace{.2 cm}
and Y. Yang\footnote{yyangc@sfu.ca}}
\date{\textit{Department of Physics, Simon Fraser University\\Burnaby, BC V5A 1S6 Canada}}
\maketitle

\thispagestyle{empty}

\vspace{2.5 cm} We study quantum corrections at the one loop level
in open superstring tachyon condensation using the boundary string
field theory (BSFT) method. We find that the tachyon field action
has the same form as at the disc level, but with a renormalized
effective coupling $\lambda' =g_s^{ren}e^{-T^2/4}$ ($g_s^{ren}$ is
the renormalized dimensionless closed string coupling,
$e^{-T^2/4}$ is the tachyonic field expectation value) and with an
effective string tension. This result is in agreement with that
based on general analysis of loop effects.

\vspace{3.8 cm} April, 2001
\end{titlepage}

\section{Introduction}
Spontaneous symmetry breaking in dual models leading to removal of
tachyonic vacuum instability has been studied some years ago
[1-4]. Recently, the problem of tachyon condensation on unstable
D-branes and on D-$\bar{\mbox{D}}$ system has been the subject of
active investigation. Two distinct methods have been used to study
this problem. The first, earlier method, uses string field theory.
The main approach here has been through Witten's cubic string
field theory \cite{witten1} and remarkable results have been
obtained using this approach \cite{2,3}. It is shown that as a
result of tachyon condensation the unstable D-brane decays into
the perturbative vacuum of closed strings or to lower dimensional
D-brane. It has been further conjectured that the tachyon
potential on a D-$\bar{\mbox{D}}$ system in type II string theory
has a universal form \cite{9911116} independent of the boundary
conformal field theory describing the D-brane, and that the open
string field theory at the tachyonic vacuum must contain closed
string states \cite{0010240}. In Witten's string field theory
approach, tachyon condensation phenomena in general involves
giving expectation values to an infinite number of components of
the string field, and as a consequence one has to resort to level
truncation \cite{witten1}.

A second approach, which appears simpler, involves using the
background independent string field theory (BSFT) of Witten
\cite{witten2,9210065} and of Shatashvili \cite{9303143}. In this
version of the open string field theory, the classical
configuration space is the space of two dimensional world sheet
theories which are conformal in the bulk (of disc topology) but
contain arbitrary boundary interaction terms. Since the tachyon
condensation is an off-shell process, the BSFT approach is well
suited to studying this problem. A large number of papers have
been written using this approach [13-21]. The BSFT method has been
extended to the case of superstrings in [17, 23-27]. The tree
level tachyonic action has been shown to be given by \be
\mathcal{S}_{\rm tree}=T_9\int d^{10}X\,e ^{-\frac{1}{4}T^2}[(2\ln
{2})
\partial_\mu T\partial^\mu T+1], \ee where $T_9$ is the tension of
the non-BPS D9-brane.

The question of quantum corrections to tachyonic action deserves a
detailed study. Partial studies of one loop effects in bosonic
case are contained in \cite{0101207,0102192}. Earlier studies on
the annulus effective action are based on the sigma model approach
to string theory \cite{tseytlin}, and also \cite{0012080,0011033},
where they calculate the Dirac-Born-Infeld (DBI) action for the
D-brane. We use the BSFT approach to calculate the tachyonic field
action in the superstring theory at one loop level. The partition
function $Z$ of the BSFT takes the form \cite{tseytlin} \be
Z&=&\sum_{\chi}g_s^{-\chi}\int d\mu(\lambda)\int e^{-I_{\rm
bulk}-I_{\rm bndy}}
[dx^\mu][d\psi][d\tilde\psi],\\
& &d\mu(\lambda)=d\lambda_1\cdots
d\lambda_N\mu(\lambda_1,\cdots,\lambda_N)\no, \ee where
$\lambda_1,\cdots,\lambda_N$ denote Teichm\"uller parameters of
the two dimensional surface with holes ($\chi=1$, $N=0$; $\chi=0$,
$N=1$; $\chi\le 1$, $N=-3\chi$). We are interested in calculating
the annulus partition function where $\chi=0$ and $N=1$. The tree
level partition function has already been evaluated; see (1).
$g_s$ denotes the unrenormalized dimensionless closed string
coupling. The measure $\mu$ in (2) includes the contribution of
the ghost determinant.

The bulk action $I_{\rm bulk}$ is given by the standard NSR action
\be I_{\rm bulk}=\frac{1}{4\pi}\int_\Sigma d^2z\left(\partial
X^\mu\bar{\partial}X_\mu
+\psi^\mu\bar{\partial}\psi_\mu+\tilde{\psi}^\mu\partial\tilde{\psi}_\mu\right),
\ee where $\psi$ and $\tilde{\psi}$ are the holomophic and
antiholomophic fermionic fields. $\Sigma$ is a two-dimensional
world sheet with the coordinates $X^\mu$. (Here we choose
$\alpha'=2$ for the closed string.)

The supersymmetric boundary action $I_{\rm bndy}$ is given by
\cite{0010108} as \be I_{\rm bndy}=\frac{1}{8\pi
}\int_{\partial\Sigma}d\theta \left[(T(X))^2+(\psi^\mu\partial_\mu
T) \frac{1}{\partial_\theta}(\psi^\nu\partial_\nu
T)+(\tilde{\psi}^\mu\partial_\mu T)
\frac{1}{\partial_\theta}(\tilde{\psi}^\nu\partial_\nu T)\right],
\ee where $T(X)$ is a tachyon profile chosen to be of the form \be
T(X)=a+u_\mu X^\mu, \ee and \be
\frac{1}{\partial_\theta}(\psi^\mu\partial_\mu T)= \frac{1}{2}\int
d\theta'\epsilon(\theta-\theta')(\psi^\mu\partial_\mu T)(\theta'),
\ee where $\epsilon(x)=+1$ for $x>0$ and $\epsilon(x)=-1$ for
$x<0$.

It was first conjectured in \cite{0010108} and later verified in
\cite{0103089,0103102} that, using Batalin-Vilkovisky (BV)
formalism \cite{batalin} of background independent superstring
field theory, the spacetime superstring field action is the
partition function defined in (2) on the disc. Since the
derivation of this result using the BV formalism appears to go
through for other geometries (the annulus here), we assume that
the string field action is given by the partition function defined
in (2) on the world sheet with the topology of an annulus. We
justify $a posteriori$ this assumption.

The plan of the paper is as follows. In section 2 we review the
results for the disc. The bulk fermionic Green function in the NS
sector is derived and the first two non-trivial terms, i.e. the
kinetic and potential energy terms, in the tachyonic action are
derived. In section 3, we first derive the required bosonic and
fermionic Green functions on the annulus. The bosonic Green
function has been derived in \cite{0101207,0102192,callan}
previously. Using these the one loop tachyonic field action is
derived. Section 4 has conclusions.

\section{Tachyon Action at the Tree Level}

In this section we choose the world sheet $\Sigma$ to be a disc
with a rotationally invariant flat metric \be
ds^2=d\sigma_1^2+d\sigma_2^2, \ee the complex variable
$z=\sigma_1+i\sigma_2$, with $|z|\leq 1$.

In (5), the constant $a$ can be shifted away by the integration
constant, so we take boundary tachyon profile of the form \be
T(X)=u_\mu X^\mu. \ee

The boundary term can then be written as \be
I'=\frac{y}{8\pi}\int_0^{2\pi}
d\theta\left(X^2+\psi\frac{1}{\partial_\theta}\psi
+\tilde{\psi}\frac{1}{\partial_\theta}\tilde{\psi}\right), \ee
where $y\equiv u^2$.

The boundary conditions are derived by varying the action
$I=I_{\rm bulk}+I_{\rm bndy}$. The bosonic Green function
satisfies, \be
\partial_z\bar\partial_{\bar z}G(z,w)=-2\pi\delta^{(2)}(z-w),
\ee
with the boundary condition \cite{9210065}
\be
z\partial_z G_B+\bar z\bar\partial_{\bar z}G_B+uG_B|_{\rho=1}=0.
\ee

The boundary conditions for the fermionic Green functions are \be
\left.\left(1+iy\frac{1}{\partial_\theta}\right)G_F\right|_{\rho=1
}
=\left.\left(1-iy\frac{1}{\partial_\theta}\right)\tilde{G}_F\right|_{\rho=1},
\ee where $G(z,w)$ and $\tilde G(\tilde z,\tilde w)$ are the
holomophic and antiholomophic Green function.

The fermionic Green functions satisfy the equations:
\be
\bar{\partial}G_F(z,w)&=&-i\sqrt{zw}\delta^{(2)} (z-w),\no\\
\partial\tilde{G}_F(\bar{z},\bar{w})&=&+i\sqrt{\bar{z}\bar{w}}\delta^{(2)} (\bar{z}-\bar{w})
\ee with the boundary conditions (12). Note that the factors
$\sqrt{zw}$ and $\sqrt{\bar{z}\bar{w}}$ before the $\delta$
function are inserted to make the Green function dimensionless.

To solve (13) with the boundary conditions (12) on the disc, we start with the ansatz,
\be
iG_F(z,w)&=&+\frac{\sqrt{zw}}{z-w}+\sum_{r}a_r (z\bar{w})^r,\no\\
i\tilde{G}_F(\bar{z},\bar{w})&=&-\frac{\sqrt{\bar{z}\bar{w}}}{\bar{z}-\bar{w}}
-\sum_{r}b_r (\bar{z}w)^r.
\ee

It is easy to verify that these are solutions of (13). $a_r$'s and
$b_r$'s are coefficients to be determined by the boundary
conditions (12).

Inserting this ansatz into the boundary conditions (12), expanding
it by series and following the same procedure as in
\cite{0101207}, we get \be
iG_F(z,w)&=&\frac{\sqrt{zw}}{z-w}-\frac{\sqrt{z\bar w}}{1-z\bar w}
+\sum_{r\geq\frac{1}{2}}\frac{2y}{r+y}(z\bar w)^r,\no\\
i\tilde{G}_F(\bar z,\bar w)&=&-\frac{\sqrt{\bar z\bar w}}{\bar z-\bar w}
+\frac{\sqrt{\bar z w}}{1-\bar z w}
-\sum_{r\geq\frac{1}{2}}\frac{2y}{r+y}(\bar z w)^r,
\ee
where the sums are taken over all positive half integers.

When $y=0$, we obtain the familiar result
\be
& &\langle\psi(\theta)\psi(\theta')\rangle|_{y=0}
+\langle\tilde{\psi}(\theta)\tilde{\psi}(\theta')\rangle|_{y=0}\no\\
&=&G_F(e^{i\theta},e^{i\theta'})|_{y=0}+\tilde{G}_F(e^{-i\theta},e^{-i\theta'})|_{y=0}\\
&=&-\frac{2}{\sin\frac{\theta-\theta'}{2}}.\no \ee

To evaluate the partition function, we need the explicit form of
the propagators of $X$ and $\psi(\tilde{\psi})$ at boundary points
$z=e^{i\theta}$ and $w=e^{i\theta'}$. That of $X$ was computed in
\cite{9210065}, \be \langle X(\theta)X(\theta')\rangle\equiv
G_B(\theta-\theta') =2\sum_{k\in
Z}\frac{1}{|k|+y}e^{ik(\theta-\theta')}. \ee

The propagator for the fermions on the boundary in the NS sector is
\be
\langle\psi(\theta)\psi(\theta')\rangle
+\langle\tilde{\psi}(\theta)\tilde{\psi}(\theta')\rangle
&\equiv&G_F(e^{i\theta},e^{i\theta'})+\tilde{G}_F(e^{-i\theta},e^{-i\theta'})\no\\
&=&2i\sum_{r\in Z+\frac{1}{2}}\frac{r}{|r|+y}e^{ir(\theta-\theta')},
\ee
which is the expected result as in \cite{0010108}.

Now we can calculate the partition function on the disc. From the
definition (2), the partition function on the disc is
($\chi=1,N=0$) \be Z=\frac{1}{g_s}\int e^{-I_{\rm bulk}-I_{\rm
bndy}}[dx^\mu][d\psi][d\tilde\psi], \ee by using the explicit form
of $I_{\rm bulk}$ and $I_{\rm bndy}$ in (3) and (4), we have \be
\frac{d}{dy}\ln Z=-\frac{1}{8\pi g_s}\int_0^{2\pi}d\theta\langle
X^2 +\psi\frac{1}{\partial_\theta}\psi
+\tilde{\psi}\frac{1}{\partial_\theta}\tilde{\psi}\rangle . \ee

The right hand side of (20) at the boundary is defined in \cite{0010108} as
\be
\langle X^2+\psi\frac{1}{\partial_\theta}\psi
+\tilde{\psi}\frac{1}{\partial_\theta}\tilde{\psi}\rangle
\equiv\lim_{\epsilon\rightarrow 0}\langle X(\theta)X(\theta+\epsilon)
+\psi(\theta)\frac{1}{\partial_\theta}\psi(\theta+\epsilon)
+\tilde{\psi}(\theta)\frac{1}{\partial_\theta}\tilde{\psi}(\theta+\epsilon)\rangle .
\ee

Define
\be
G'_F(\epsilon;y)&\equiv&
\langle\psi(\theta)\frac{1}{\partial_\theta}\psi(\theta+\epsilon)\rangle,\no\\
\tilde{G}'_F(\epsilon;y)&\equiv&
\langle\tilde{\psi}(\theta)\frac{1}{\partial_\theta}\tilde{\psi}
(\theta+\epsilon)\rangle . \ee

We obtain
\be
G'_F(\epsilon;y)+\tilde{G}'_F(\epsilon;y)
&=&-2\sum_{r\in Z+\frac{1}{2}}\frac{1}{|r|+y}e^{ir\epsilon}\no\\
&=&G_B(\epsilon;y)-2G_B(\frac{\epsilon}{2};2y).
\ee

This gives \be \langle X^2+\psi\frac{1}{\partial_\theta}\psi
+\tilde{\psi}\frac{1}{\partial_\theta}\tilde{\psi}\rangle=-8\ln
{2}+2f(y)-2f(2y), \ee where \be
f(y)=\frac{2}{y}-4y\sum_{k=1}^{\infty}\frac{1}{k(k+y)}. \ee

Integrating over $y$, we find that the partition function is given by
\be
Z(y)=Z'4^y\frac{Z_B(y)^2}{Z_B(2y)},
\ee
where $Z'$ is an integration constant and $Z_B$ is the partition function for the bosonic
case \cite{9210065} given by,
\be
Z_B(y)=\sqrt{y}e^{\gamma y}\Gamma(y).
\ee

According to the conjecture we mentioned in the introduction, for
the superstring theory, the BSFT action is simply this partition
function. Writing down the action in terms of tachyonic field
$T=uX$, we find the action of the tachyonic field as \be
\mathcal{S}_{tree}=\frac{T_9}{g_s}\int
d^{10}X\,e^{-\frac{1}{4}T^2}((2\ln {2})\partial_\mu T\partial^\mu
T+1) \ee where $T_9$ is the tension of the non-BPS D9-brane. The
integration constant $Z'$ can then be fixed by comparing to the
standard tension of a non-BPS D9-brane \cite{0003101,9909072}.

\section{Tachyon Action at the One Loop level}

In this section, we choose the world sheet $\Sigma$ to be an
annulus with a rotationally invariant flat metric \be
ds^2=d\sigma_1^2+d\sigma_2^2, \ee the complex variable
$z=\sigma_1+i\sigma_2$, with $a\leq |z|\leq b$.

We first take the constant tachyon profile $T(X)=a$, Then the
direct calculation gives\be \mathcal{S}_{\rm
1-loop}=V_0e^{-\frac{1}{2}T^2}. \ee Thus the tachyon potential of
the one-loop level is \be V(T)\sim e^{-\frac{1}{2}T^2}. \ee
 Next, we take the
tachyon profile as in (8), \be T(X)=\left\{\ba{rr} u_aX&\rho=a\\
u_bX&\rho=b \ea\right., \ee and the boundary term can be written
as \be I'=\frac{y_b}{8\pi}\int_0^{2\pi}d\theta
\left(X^2+\psi\frac{1}{\partial_\theta}\psi
+\tilde{\psi}\frac{1}{\partial_\theta}\tilde{\psi}\right)_{\rho=b}+\frac{y_a}{8\pi}\int_0^{2\pi}
d\theta\left(X^2+\psi\frac{1}{\partial_\theta}\psi
+\tilde{\psi}\frac{1}{\partial_\theta}\tilde{\psi}\right)_{\rho=a},
\ee where $y_a\equiv u_a^2$ and $y_b\equiv u_b^2$.

The boundary conditions for the Green function of bosonic field are \cite{0101207,0102192}
\be
(z\partial +\bar z\bar\partial -y_a)G_B(z,w)|_{\rho=a}&=&0,\no\\
(z\partial +\bar z\bar\partial +y_b)G_B(z,w)|_{\rho=b}&=&0.
\ee

To solve for $G_B(z,w)$, we start with the ansatz,
\be
G_B(z,w)&=&-\ln |z-w|^2+C_1\ln |z|^2\ln |w|^2+C_2(\ln |z|^2+\ln |w|^2)+C_3\no\\
& &+\sum_{-\infty}^{\infty}a_k[(z\bar w)^k+(\bar z w)^k]
+\sum_{-\infty}^{\infty}b_k\left[\left(\frac{z}{w}\right)^k
+\left(\frac{\bar z}{\bar w}\right)^k\right]
\ee

As before, one can easily verify that these are solutions of (10)
on an annulus. $C_1$, $C_2$, $C_3$, $a_r$'s and $b_r$'s are
coefficients to be determined by the boundary conditions (34).

Inserting this ansatz into the boundary conditions (34), expanding
it by series and following the procedure as in \cite{0101207}, we
get

\be
& &G_B(z,w)\no\\
&=&-\ln (|z-w|^2/b^2)\no\\
& &-\frac{[y_a\ln ({|z|^2}/{b^2})+2][y_b\ln ({|w|^2}/{b^2})-2]
-2(y_b+y_a)\ln (|w|^2/b^2)+2y_a\ln ({a^2}/{b^2})}{2y_a+2y_b-y_ay_b\ln (a^2/b^2)}\no\\
& &-\sum_{n=1}^{\infty}\ln
\left[\left|1-\left(\frac{a}{b}\right)^{2n}\frac{z\bar
w}{a^2}\right|^2
\cdot\left|1-\left(\frac{a}{b}\right)^{2n}\frac{b^2}{z\bar
w}\right|^2
\cdot\left|1-\left(\frac{a}{b}\right)^{2n}\frac{z}{w}\right|^2
\cdot\left|1-\left(\frac{a}{b}\right)^{2n}\frac{w}{z}\right|^2\right]\no\\
& &-2\sum_{k=1}^{\infty}\frac{[y_b(k+y_a)b^{2k}+y_a(k-y_b)a^{2k}]a^{2k}}
{k(b^{2k}-a^{2k})[(k+y_a)(k+y_b)b^{2k}-(k-y_a)(k-y_b)a^{2k}]}
\left[\left(\frac{z\bar{w}}{a^2}\right)^k+\left(\frac{\bar z w}{a^2}\right)^k\right]\no\\
& &-2\sum_{k=1}^{\infty}\frac{[y_a(k+y_b)b^{2k}-y_b(k-y_a)a^{2k}]a^{2k}}
{k(b^{2k}-a^{2k})[(k+y_a)(k+y_b)b^{2k}-(k-y_a)(k-y_b)a^{2k}]}
\left[\left(\frac{b^2}{z\bar w}\right)^k+\left(\frac{b^2}{\bar z w}\right)^k\right]\no\\
& &-2\sum_{k=1}^{\infty}\frac{(y_b+y_a)ka^{2k}b^{2k}}
{k(b^{2k}-a^{2k})[(k+y_a)(k+y_b)b^{2k}-(k-y_a)(k-y_b)a^{2k}]}
\left[\left(\frac{z}{w}\right)^k+\left(\frac{\bar z}{\bar w}\right)^k\right]\no\\
& &-2\sum_{k=1}^{\infty}\frac{(y_b+y_a)ka^{2k}b^{2k}}
{k(b^{2k}-a^{2k})[(k+y_a)(k+y_b)b^{2k}-(k-y_a)(k-y_b)a^{2k}]}
\left[\left(\frac{w}{z}\right)^k+\left(\frac{\bar w}{\bar z}\right)^k\right]
\ee

It is readily seen that the above boundary conditions (34) are
consistent with the Gauss' law constraint \be
\int_{\rho=a,b}\left(\frac{\partial G}{\partial
n}\right)ds=-2\pi\alpha'\mbox{\hspace{1 cm}(here $\alpha'=2$ for
closed string.)} \ee

To see this, note that (36) contains all the terms in the solution
in \cite{callan} plus additional $y$ dependent terms. These
additional terms yield zero contribution to the integral above.

The boundary conditions for the Green function of the fermionic field are
\be
\left.\left(1-iy_a\frac{1}{\partial_\theta}\right)G_F\right|_{\rho=a}
&=&\left.\left(1+iy_a\frac{1}{\partial_\theta}\right)\tilde{G}_F\right|_{\rho=a},\no\\
\left.\left(1+iy_b\frac{1}{\partial_\theta}\right)G_F\right|_{\rho=b}
&=&\left.\left(1-iy_b\frac{1}{\partial_\theta}\right)\tilde{G}_F\right|_{\rho=b}.
\ee

To solve the eq. (13) with the boundary conditions (38) on the
annulus, we start with the ansatz, \be
iG_F(z,w)&=&\frac{\sqrt{zw}}{z-w} +\sum_{r\in Z+\frac{1}{2}}a_r
(z\bar{w})^r +\sum_{r\in Z+\frac{1}{2}}a'_r \left(\frac{z}{w}\right)^r,\no\\
i\tilde{G}_F(\bar{z},\bar{w})&=&-\frac{\sqrt{\bar{z}\bar{w}}}{\bar{z}-\bar{w}}
-\sum_{r\in Z+\frac{1}{2}}b_r (\bar{z}w)^r -\sum_{r\in
Z+\frac{1}{2}}b'_r \left(\frac{\bar{z}}{\bar{w}}\right)^r. \ee

After a straightforward, albeit lengthy, calculation we get the
following results for $G_F(z,w)$ and $\tilde{G}_F(\bar z,\bar w)$
\be iG_F(z,w)
&=&\frac{\sqrt{zw}}{z-w}\no\\
& &-\sum_{n=1}^{\infty}\frac{\sqrt{\left(\frac{a}{b}\right)^{2n}\frac{z\bar w}{a^2}}}
{1-\left(\frac{a}{b}\right)^{2n}\frac{z\bar w}{a^2}}
+\sum_{n=1}^{\infty}\frac{\sqrt{\left(\frac{a}{b}\right)^{2n}\frac{b^2}{z\bar w}}}
{1-\left(\frac{a}{b}\right)^{2n}\frac{b^2}{z\bar w}}\no\\
& &-\sum_{n=1}^{\infty}\frac{\sqrt{\left(\frac{a}{b}\right)^{2n}\frac{z}{w}}}
{1-\left(\frac{a}{b}\right)^{2n}\frac{z}{w}}
2\sum_{n=1}^{\infty}\frac{\sqrt{\left(\frac{a}{b}\right)^{2n}\frac{w}{z}}}
{1-\left(\frac{a}{b}\right)^{2n}\frac{w}{z}}\no\\
& &+2\sum_{r\geq\frac{1}{2}}\frac{[y_b(r+y_a)b^{2r}+y_a(r-y_b)a^{2r}]a^{2r}}
{(b^{2r}-a^{2r})[(r+y_a)(r+y_b)b^{2r}-(r-y_a)(r-y_b)a^{2r}]}
\left(\frac{z\bar{w}}{a^2}\right)^r\no\\
& &-2\sum_{r\geq\frac{1}{2}}\frac{[y_a(r+y_b)b^{2r}+y_b(r-y_a)a^{2r}]a^{2r}}
{(b^{2r}-a^{2r})[(r+y_a)(r+y_b)b^{2r}-(r-y_a)(r-y_b)a^{2r}]}
\left(\frac{b^2}{z\bar{w}}\right)^r\no\\
& &+2\sum_{r\geq\frac{1}{2}}\frac{(y_b+y_a)ra^{2r}b^{2r}}
{(b^{2r}-a^{2r})[(r+y_a)(r+y_b)b^{2r}-(r-y_a)(r-y_b)a^{2r}]}
\left(\frac{z}{w}\right)^r\no\\
& &-2\sum_{r\geq\frac{1}{2}}\frac{(y_b+y_a)ra^{2r}b^{2r}}
{(b^{2r}-a^{2r})[(r+y_a)(r+y_b)b^{2r}-(r-y_a)(r-y_b)a^{2r}]}
\left(\frac{w}{z}\right)^r,
\ee
and
\be
i\tilde{G}_F(z,w)&=&-\frac{\sqrt{\bar z\bar w}}{\bar z-\bar w}\no\\
& &+\sum_{n=1}^{\infty}\frac{\sqrt{\left(\frac{a}{b}\right)^{2n}\frac{\bar z w}{a^2}}}
{1-\left(\frac{a}{b}\right)^{2n}\frac{\bar z w}{a^2}}
-\sum_{n=1}^{\infty}\frac{\sqrt{\left(\frac{a}{b}\right)^{2n}\frac{b^2}{\bar z w}}}
{1-\left(\frac{a}{b}\right)^{2n}\frac{b^2}{\bar z w}}\no\\
& &+\sum_{n=1}^{\infty}\frac{\sqrt{\left(\frac{a}{b}\right)^{2n}\frac{\bar z}{\bar w}}}
{1-\left(\frac{a}{b}\right)^{2n}\frac{\bar z}{\bar w}}
-\sum_{n=1}^{\infty}\frac{\sqrt{\left(\frac{a}{b}\right)^{2n}\frac{b^2}{\bar w\bar z}}}
{1-\left(\frac{a}{b}\right)^{2n}\frac{\bar w}{\bar z}}\no\\
& &-2\sum_{r\geq\frac{1}{2}}\frac{[y_b(r+y_a)b^{2r}+y_a(r-y_b)a^{2r}]a^{2r}}
{(b^{2r}-a^{2r})[(r+y_a)(r+y_b)b^{2r}-(r-y_a)(r-y_b)a^{2r}]}
\left(\frac{\bar z w}{a^2}\right)^r\no\\
& &+2\sum_{r\geq\frac{1}{2}}\frac{[y_a(r+y_b)b^{2r}+y_b(r-y_a)a^{2r}]a^{2r}}
{(b^{2r}-a^{2r})[(r+y_a)(r+y_b)b^{2r}-(r-y_a)(r-y_b)a^{2r}]}
\left(\frac{b^2}{\bar z w}\right)^r\no\\
& &-2\sum_{r\geq\frac{1}{2}}\frac{(y_b+y_a)ra^{2r}b^{2r}}
{(b^{2r}-a^{2r})[(r+y_a)(r+y_b)b^{2r}-(r-y_a)(r-y_b)a^{2r}]}
\left(\frac{\bar z}{\bar w}\right)^r\no\\
& &+2\sum_{r\geq\frac{1}{2}}\frac{(y_b+y_a)ra^{2r}b^{2r}}
{(b^{2r}-a^{2r})[(r+y_a)(r+y_b)b^{2r}-(r-y_a)(r-y_b)a^{2r}]}
\left(\frac{\bar w}{\bar z}\right)^r.
\ee

To evaluate the partition function, we need to calculate the
propagators on the boundary, for both the bosonic and the
fermionic fields. We first work at the $\rho=b$ boundary. Let
$z=be^{i\theta}$ and $w=be^{i(\theta+\epsilon)}$, we have \be
& &\lim_{\epsilon\rightarrow 0}G_B(y_a,y_b;\epsilon;a,b)\no\\
&\equiv&\lim_{\epsilon\rightarrow 0}G_B(be^{i\theta},be^{i(\theta+\epsilon)})\no\\
&=&-2\ln (1-e^{i\epsilon})-2\ln (1-e^{-i\epsilon})
-8\sum_{n=1}^{\infty}\ln
\left[1-\left(\frac{a}{b}\right)^{2n}\right]
+\frac{4-2y_a\ln \frac{a^2}{b^2}}{2y_a+2y_b-y_ay_b\ln \frac{a^2}{b^2}}\no\\
& &-4\sum_{k=1}^{\infty}\frac{[y_b(k+y_a)b^{2k}+y_a(k-y_b)a^{2k}]b^{2k}}
{k(b^{2k}-a^{2k})[(k+y_a)(k+y_b)b^{2k}-(k-y_a)(k-y_b)a^{2k}]}\no\\
& &-4\sum_{k=1}^{\infty}\frac{[y_a(k+y_b)b^{2k}+y_b(k-y_a)a^{2k}]a^{2k}}
{k(b^{2k}-a^{2k})[(k+y_a)(k+y_b)b^{2k}-(k-y_a)(k-y_b)a^{2k}]}\no\\
& &-8\sum_{k=1}^{\infty}\frac{(y_b+y_a)ka^{2k}b^{2k}}
{k(b^{2k}-a^{2k})[(k+y_a)(k+y_b)b^{2k}-(k-y_a)(k-y_b)a^{2k}]}\no\\
&\equiv&-2\ln (1-e^{i\epsilon})-2\ln
(1-e^{-i\epsilon})+F(y_a,y_b;a,b), \ee where $F(y_a,y_b;a,b)\equiv
F(y_a,y_b;\epsilon=0;a,b)$ are the terms which are not singular
when we take the limit $\epsilon=0$.

As in the case of disc topology, we define
\be
G'_F(\epsilon,y_a,y_b)&\equiv&G'_F(be^{+i\theta},be^{+i(\theta+\epsilon)})
=\langle\psi(\theta)\frac{1}{\partial_\theta}\psi(\theta+\epsilon)\rangle ,\no\\
\tilde{G}'_F(\epsilon,y_a,y_b)&\equiv&\tilde{G}'_F(be^{-i\theta},be^{-i(\theta+\epsilon)})
=\langle\tilde{\psi}(\theta)\frac{1}{\partial_\theta}\tilde{\psi}(\theta+\epsilon)\rangle .
\ee

We expand $G'_F(\epsilon,y_a,y_b)$ and
$\tilde{G}'_F(\epsilon,y_a,y_b)$, then compare them to the
expansion of $G_B(y_a,y_b;\epsilon;a,b)$. By a straightforward
calculation it can be shown that \be
& &G'_F(\epsilon,y_a,y_b)+\tilde{G}'_F(\epsilon,y_a,y_b)\no\\
&=&G_B(y_a,y_b;\epsilon;a,b)-2G_B(2y_a,2y_b;\frac{\epsilon}{2};\sqrt{a},\sqrt{b}).
\ee

Now, we are ready to calculate the right hand side of (20) at the
boundary $\rho=b$. From eqn. (21), we have \be & &\langle
X^2(\theta) +\psi(\theta)\frac{1}{\partial_\theta}\psi(\theta)
+\tilde{\psi}(\theta)\frac{1}{\partial_\theta}\tilde{\psi}(\theta)\rangle|_{\rho=b}\no\\
&=&\lim_{\epsilon\rightarrow 0}[G_B(y_a,y_b;\epsilon;a,b)
+G'_F(\epsilon,y_a,y_b)+\tilde{G}'_F(\epsilon,y_a,y_b)]\no\\
&=&2\lim_{\epsilon\rightarrow 0}[G_B(y_a,y_b;\epsilon;a,b)
-G_B(2y_a,2y_b;\frac{\epsilon}{2};\sqrt{a},\sqrt{b})]\no\\
&=&-8\ln 2+2F(y_a,y_b;a,b)-2F(2y_a,2y_b;\sqrt{a},\sqrt{b}). \ee

From its definition (2), the partition function on an annulus
($\chi=0$, and $N=1$) is \be Z=\int d\lambda\mu(\lambda)\int
e^{-I_{\rm bulk}-I_{\rm
bndy}}[dx^\mu][d\psi][d\tilde\psi]\equiv\int d\lambda\, Z(a/b),
\ee where $\lambda =\ln (a/b)$ is the Teichm\"uller parameter on
the annulus. $\mu(\lambda)$ includes the contribution of the ghost
determinant. In superstring theory, the ghost contribution of the
bosonic fields exactly cancels out that of the fermionic fields at
the one loop level. So in our case (annulus), $\mu(\lambda)=1$.

Differentiating with respect to the parameter $y_b$, we have \be
\frac{\partial}{\partial y_b}\ln
Z(a/b)&=&-\frac{1}{8\pi}\int_0^{2\pi} d\theta\langle X^2(\theta)
+\psi(\theta)\frac{1}{\partial_\theta}\psi(\theta)
+\tilde{\psi}(\theta)\frac{1}{\partial_\theta}\tilde{\psi}(\theta)\rangle|_{\rho=b}\no\\
&=&(2\ln
{2})-\frac{1}{2}F(y_a,y_b;a,b)+\frac{1}{2}F(2y_a,2y_b;\sqrt{a},\sqrt{b}).
\ee

Integrating over $y_b$, up to an integration constant, we get \be
\ln Z(a/b)&=&(2\ln 2)y_b
-\frac{1}{2}\ln \left[y_a+y_b-\frac{y_ay_b}{2}\ln \left(\frac{a^2}{b^2}\right)\right]\no\\
& &+\sum_{k=1}^\infty \left\{\ln
\left[\left(1+\frac{2y_a}{k}\right)\left(1+\frac{2y_b}{k}\right)
-\left(1-\frac{2y_a}{k}\right)\left(1-\frac{2y_b}{k}\right)
\left(\frac{a}{b}\right)^k\right]\right.\no\\
& &\left.-2\ln
\left[(1+\frac{y_a}{k})\left(1+\frac{y_b}{k}\right)-\left(1-\frac{y_a}{k}\right)
\left(1-\frac{y_b}{k}\right)
\left(\frac{a}{b}\right)^{2k}\right]\right\} +f(y_a), \ee where
$f(y_a)$ is an arbitrary function of $y_a$.

Similarly, we can compute the propagator at the boundary $\rho=a$.
Repeating the above procedure, we obtain \be \ln Z(a/b)&=&(2\ln
2)y_a
-\frac{1}{2}\ln \left[y_a+y_b-\frac{y_ay_b}{2}\ln \left(\frac{a^2}{b^2}\right)\right]\no\\
& &+\sum_{k=1}^\infty \left\{\ln
\left[\left(1+\frac{2y_a}{k}\right)\left(1+\frac{2y_b}{k}\right)
-\left(1-\frac{2y_a}{k}\right)\left(1-\frac{2y_b}{k}\right)
\left(\frac{a}{b}\right)^k\right]\right.\no\\
& &\left.-2\ln
\left[\left(1+\frac{y_a}{k}\right)\left(1+\frac{y_b}{k}\right)
-\left(1-\frac{y_a}{k}\right)\left(1-\frac{y_b}{k}\right)
\left(\frac{a}{b}\right)^{2k}\right]\right\} +f(y_b), \ee where
$f(y_b)$ is an arbitrary function of $y_b$.

Comparing the results (48) and (49), we can fix the arbitrary
functions $f(y_a)$ and $f(y_b)$. The final expression is \be \ln
Z(a/b)&=&(2\ln 2)y_b+(2\ln 2)y_a
-\frac{1}{2}\ln \left[y_a+y_b-\frac{y_ay_b}{2}\ln \left(\frac{a^2}{b^2}\right)\right]\no\\
& &+\sum_{k=1}^\infty \left\{\ln
\left[\left(1+\frac{2y_a}{k}\right)\left(1+\frac{2y_b}{k}\right)
-\left(1-\frac{2y_a}{k}\right)\left(1-\frac{2y_b}{k}\right)
\left(\frac{a}{b}\right)^k\right]\right.\no\\
& &\left.-2\ln
\left[\left(1+\frac{y_a}{k}\right)\left(1+\frac{y_b}{k}\right)
-\left(1-\frac{y_a}{k}\right)\left(1-\frac{y_b}{k}\right)
\left(\frac{a}{b}\right)^{2k}\right]\right\}. \ee

The partition function, then, can be obtained as \be
Z(a/b)=Z'4^{y_b+y_a}\frac{Z^2_B(y_a,y_b;a,b)}{Z_B(2y_a,2y_b;\sqrt{a},\sqrt{b})},
\ee where $Z'$ is the integration constant which we can choose to
be the same as the disc case for convenience, and \be
Z_B(y_a,y_b;a,b)&=&\left[y_a+y_b-\frac{y_ay_b}{2} \ln
\left(\frac{a^2}{b^2}\right)\right]^{-1/2}\no\\ &
&\cdot\prod_{k=1}^\infty\left[(1+\frac{y_a}{k})(1+\frac{y_b}{k})
-(1-\frac{y_a}{k})(1-\frac{y_b}{k})\left(\frac{a}{b}\right)^{2k}\right]^{-1}
\ee is the bosonic partition function on the annulus.

It is convenient to take the outer radius $b=1$ in the following.
Integrating over the modulus $d\lambda=da/a$, we obtain \be
Z&=&\int d\lambda Z(a)\no\\
&=&\sqrt{2}Z'4^{y_b+y_a}\int_0^1\frac{da}{a}\left[y_a+y_b-\frac{y_ay_b}{2}
\ln {a^2}\right]^{-1/2}\no\\ &
&\cdot\prod_{k=1}^\infty\frac{(1+\frac{2y_a}{k})(1+\frac{2y_b}{k})
-(1-\frac{2y_a}{k})(1-\frac{2y_b}{k})a^k}
{\left[(1+\frac{y_a}{k})(1+\frac{y_b}{k})-(1-\frac{y_a}{k})(1-\frac{y_b}{k})
a^{2k}\right]^2}. \ee

In the one loop level, there are two types of corrections coming
from two different configurations:

\begin{description}

\item[Case 1.]

The two ends of the open strings end on two different D-branes on
which they may have different tachyonic couplings $y_a$ and $y_b$.
In this case, as above, $Z$ is a function of both $y_a$ and $y_b$.
To write the background independent action in term of the
tachyonic field on one of the two D-branes, say $\rho=b$ boundary,
we take $y_a=0$ and $y_b=y$, then \be
Z=Z'4^y\int_0^1\frac{da}{a}\sqrt{\frac{2}{y}}
\cdot\prod_{k=1}^\infty\frac{(1+\frac{2y}{k})-(1-\frac{2y}{k})a^k}
{\left[(1+\frac{y}{k})-(1-\frac{y}{k})a^{2k}\right]^2} \ee

Expanding (54) in powers of $y$, the expression of $Z$ takes the
form \be
Z&=&Z'\int_0^1\frac{da}{a}\prod_{k=1}^\infty\left(\frac{1-a^{2k-1}}{1-a^{2k}}\right)\no\\
& &\cdot\left(\sqrt{\frac{2}{y}}+\left[2\sqrt{2}\ln {2}
-4\sqrt{2}\sum_{n=1}^\infty\ln (1-a^{2n-1})\right]
\sqrt{y}+\cdots\right). \ee

Writing the first two terms (corresponding to the kinetic term and
the tachyon potential) of the partition function in terms of the
tachyonic field, we obtain the tachyon action \be \mathcal{S}_{\rm
1-loop}&=&T_9\int_0^1\frac{da}{a}
\prod_{k=1}^\infty\left(\frac{1-a^{2k-1}}{1-a^{2k}}\right)\no\\
& &\int d^{10}X\,e^{-\frac{1}{4}T^2}\left(\left[2\ln {2}
-4\sum_{n=1}^\infty\ln (1-a^{2n-1})\right]\partial_\mu
T\partial^\mu T+1\right). \ee

Let us consider the integral over the modulus $a$. The integral
$\int_0^1\frac{da}{a}\prod_{k=1}^\infty\left(\frac{1-a^{2k-1}}{1-a^{2k}}\right)$
is divergent at $a=0$. Introducing a cut-off parameter $\Lambda$,
we may write this integral as \be
\int_0^1\frac{da}{a}\prod_{k=1}^\infty\left(\frac{1-a^{2k-1}}{1-a^{2k}}\right)
=\Lambda +\Lambda_{\rm finite}, \ee where $\Lambda_{\rm finite}$
is the finite part of the integral. The cut off $\Lambda$ will be
absorbed by renormalization.

In the coefficient of the kinetic energy term, the integrand \be
-4\int_0^1\frac{da}{a}\prod_{k=1}^\infty\left(\frac{1-a^{2k-1}}{1-a^{2k}}\right)
\sum_{n=1}^\infty\ln (1-a^{2n-1})\ee is finite, so we can ignore
this term comparing to the first term of the kinetic energy term.
Thus the one-loop action can be written as \be \mathcal{S}_{\rm
1-loop}&=&T_9(\Lambda+\Lambda_{\rm finite})\int
d^{10}X\,e^{-\frac{1}{4}T^2}\left[(2\ln {2})\partial_\mu
T\partial^\mu T+1\right]. \ee

\item[Case 2.]

The two ends of open strings end on different D-branes, but with
the same coupling $y_a=y_b\equiv y$. In this case, the partition
function $Z(y_a, y_b)|_{y_a=y_b\equiv y}$ may be used, since the
ends are independent. The partition function is given
by\footnote{Note the factor before the integration is $4^y$, not
$4^{2y}$ as it might appear to do so from (52). This is because
when we set $y_a=y_b\equiv y$, we do not need to add the extra
terms $f(y_a)$ and $f(y_b)$ in (46) and (47).}, \be
Z&=&Z'4^{y}\int_0^1\frac{da}{a}\sqrt{\frac{1}{y-(y^2/4)\ln {a^2}}}
\cdot\prod_{k=1}^\infty\frac{(1+\frac{2y}{k})^2-(1-\frac{2y}{k})^2a^k}
{\left[(1+\frac{y}{k})^2-(1-\frac{y}{k})^2a^{2k}\right]^2}\ee

Again expanding in $y$, we obtain \be
Z&=&Z'\int_0^1\frac{da}{a}\sqrt{\frac{1}{1-(y/2)\ln {a}}}
\cdot\prod_{k=1}^\infty\left(\frac{1-a^{2k-1}}{1-a^{2k}}\right)\no\\
& & \cdot\left(\sqrt{\frac{1}{y}}+\left[2\ln
{2}-8\sum_{n=1}^\infty\ln (1-a^{2n-1})\right]
\sqrt{y}+\cdots\right). \ee Note that we keep the square root form
of the zero mode without expanding it in $y$, this is because that
it is not proper to expand it for small $y$ at the integral end
$a=0$.

The integral in (61) has a divergence at $a=0$. (Similarly, here
we can ignore the second term, which is finite, of the kinetic
energy term.) This is handled by replaces the lower limit $a=0$ by
$a=\delta$ and lets $\delta\rightarrow 0$ at the end. Consider the
integral \be I&=&\lim_{\delta\rightarrow
0}\int_\delta^1\frac{da}{a}\frac{1}{\sqrt{1-(y/2)\ln (a)}}\prod_{k=1}^\infty\left(\frac{1-a^{2k-1}}{1-a^{2k}}\right)\no\\
&=&\lim_{\delta\rightarrow
0}\int_\delta^1\frac{da}{a}\frac{1}{\sqrt{1-(y/2)\ln
(a)}}\left[1+\sum_{n=1}^\infty C_n a^n\right],\ee where $C_n$ are
coefficients whose form is not needed in what follows, but
$\sum_{n=1}^\infty C_n a^n$ is convergent.

This integral can be done and the result is\be
I=\lim_{\delta\rightarrow
0}-\frac{4}{y}\left(1-\sqrt{1-\frac{y}{2}\ln \delta
}\right)+\mbox{finite terms}\sim\frac{\ln \delta }{\sqrt{y}}, \ee
therefore, \be Z\sim \left(\frac{1}{y}+2\ln {2}\right). \ee we
obtain the tachyon action in the form \be \mathcal{S}_{\rm
1-loop}\sim\int d^{10}X\,e^{-\frac{1}{2}T^2\partial_\mu
T\partial^\mu T}[(2\ln {2} )\partial_\mu T\partial^\mu T+1]. \ee

\item[Case 3.]

The form of tachyon action in case 2 is unexpected according to
our early analysis (31). The reason is that we set $y_a=y_b$ after
we integrated over $y_b$. In fact, this is the case which
corresponds to a open string ending on two different branes.
Another choice is to set $y_a=y_b=y$ before integrating over
$y_b$. Since the ends are not independent in this case, $Z$ is
only a function of $y$, and thus this corresponds to an open
string ending on only a single brane. This should be the proper
correction to the disc case. To do this, let's go back to equation
(47) and set $y_a=y_b\equiv y$ there; we obtain\be
\frac{\partial}{\partial y}\ln
Z(a/b)&=&-\frac{1}{8\pi}\int_0^{2\pi} d\theta\langle X^2(\theta)
+\psi(\theta)\frac{1}{\partial_\theta}\psi(\theta)
+\tilde{\psi}(\theta)\frac{1}{\partial_\theta}\tilde{\psi}(\theta)\rangle|_{\rho=b}\no\\
&=&(2\ln
{2})-\frac{1}{2}F(y,y;a,b)+\frac{1}{2}F(2y,2y;\sqrt{a},\sqrt{b}).
\ee

Integrating over $y$, up to an integration constant, we get \be
\ln Z(a/b)&=&(2\ln 2)y
-\frac{1}{4}\ln \left[2y-y^2\ln \left(\frac{a}{b}\right)\right]\no\\
& &+\frac{1}{2}\sum_{k=1}^\infty \left\{\ln
\left[\left(1+\frac{2y}{k}\right)^2-\left(1-\frac{2y}{k}\right)^2
\left(\frac{a}{b}\right)^k\right]\right.\no\\
& &\left.-2\ln
\left[\left(1+\frac{y}{k}\right)^2-\left(1-\frac{y}{k}\right)^2
\left(\frac{a}{b}\right)^{2k}\right]\right\}, \ee

Taking the outer radius $b=1$ in the following. Integrating over
the modulus $d\lambda=da/a$, we obtain \be
Z&=&\int d\lambda Z(a)\no\\
&=&\sqrt{2}Z'4^y\int_0^1\frac{da}{a}\left(2y-y^2 \ln
{a}\right)^{-1/4}\cdot\prod_{k=1}^\infty\left[\frac{(1+\frac{2y}{k})^2
-(1-\frac{2y}{k})^2}{\left[(1+\frac{y}{k})^2-(1-\frac{y}{k})^2
a^{2k}\right]^2}\right]^{1/2}\no\\
&=&\sqrt{2}Z'\int_0^1\frac{da}{a}\left[2y-y^2 \ln
{a}\right]^{-1/4}
\cdot\prod_{k=1}^\infty\left(\frac{1-a^{2k-1}}{1-a^{2k}}\right)^{1/2}\no\\
& & \cdot\left(1+\left[2\ln {2}-4\sum_{n=1}^\infty\ln
(1-a^{2n-1})\right] y+\cdots\right). \ee

We handle the integral over $a$ by the same way as in case 2. We
get\be I&=&\lim_{\delta\rightarrow
0}\int_\delta^1\frac{da}{a}\left[2y-y^2 \ln {a}\right]^{-1/4}
\cdot\prod_{k=1}^\infty\left(\frac{1-a^{2k-1}}{1-a^{2k}}\right)^{1/2}\no\\
&=&\lim_{\delta\rightarrow
0}-\frac{4}{3y^2}\left((2y)^{3/4}-(2y-y^2)\ln \delta
)^{3/4}\right)+\mbox{finite terms}\no\\
&\sim&\frac{\ln \delta}{\sqrt{y}} \ee Retaining the first two
terms of the partition function, \be Z\sim
Z'\left(\frac{1}{\sqrt{y}}+2\ln {2}\sqrt{y}\right)\cdot\ln
\delta\equiv(\Lambda+\Lambda_{\rm
finite})Z'\left(\frac{1}{\sqrt{y}}+2\ln {2}\sqrt{y}\right). \ee we
obtain the tachyon action in the form \be \mathcal{S}_{\rm
1-loop}=T_9(\Lambda+\Lambda_{\rm finite})\int
d^{10}X\,e^{-\frac{1}{2}T^2}[(2\ln {2} )\partial_\mu T\partial^\mu
T+1]. \ee
\end{description}

When we consider the D9-brane, which is actually the whole
10-dimensional spacetime, both ends of open strings must end on
the same D9-brane. Thus we should use the one loop tachyon action
of case 3 as the quantum correction of the tree level action. By
the definition of the effective action of tachyonic fields (2), we
get \be \mathcal{S}&=&\mathcal{S}_{\rm tree}+\mathcal{S}_{\rm 1-loop}\no\\
&=&\frac{T_9}{g_s}\int d^{10}X\,
e^{-\frac{1}{4}T^2}[(2\ln {2})\partial_\mu T\partial^\mu T+1]\no\\
& &+T_9(\Lambda+\Lambda_{\rm finite})\int
d^{10}X\,e^{-\frac{1}{2}T^2}[(2\ln {2} )\partial_\mu T\partial^\mu
T+1]\no\\ &=&T_9\int d^{10}X\,e^{-\frac{1}{2}T^2}\left[
\left(\frac{1}{\lambda}+\Lambda +\Lambda_{\rm finite}\right)[(2\ln
{2} )\partial_\mu T\partial^\mu T+1]\right], \ee where we define
\cite{0009148} \be \lambda=g_se^{-\frac{1}{4}T^2} \ee as the
effective string coupling. The infinite cut-off $\Lambda$ can be
absorbed into the renormalized effective string coupling
$\lambda'$ defined by \be
\frac{1}{\lambda'}=\frac{1}{\lambda}+\Lambda. \ee

Thus we can write the tachyon action up to the one loop quantum
correction as \be \mathcal{S}&=&\frac{T'_9}{\lambda'}\int
d^{10}X\,e^{-\frac{1}{2}T^2}[(2\ln {2})\partial_\mu T\partial^\mu
T+1], \ee where the one loop corrected D9-brane tension is (in
terms of $\alpha' =2$) \be T'_9=(1+\lambda')T_9 \ee and the one
loop quantum corrected tachyon potential \be
V(T)=e^{-\frac{1}{2}T^2}\ee which is the square of the tree level
tachyon potential $e^{-\frac{1}{4}T^2}$.

\section{Conclusions}
We have studied quantum effects for tachyon condensation using
background independent string field theory methods. In particular,
the contribution to open superstring partition function from the
annulus diagram is calculated. We have shown that there are three
important contributions to the tachyonic field action at one loop
level . First, the tachyonic potential is proportional to
$e^{-\frac{1}{2}T^2}$ and not $e^{-\frac{1}{4}T^2}$ as for the
tree level. The second point is that there arises a
renormalization of the effective string coupling given by
$\lambda' =g_s^{ren}e^{-T^2/4}$, where $g_s^{ren}$ is the
renormalized closed string coupling constant defined by (73) and
(74). We note that this involves a field dependent
renormalization. Finally, there is a finite one-loop contribution
to the D9-brane tension given by (76).

Except for the renormalized quantities $\lambda'$, $T'_9$ and
$e^{-T^2/2}$, the one loop corrected tachyon action has the same
form as the tree level one, which is expected based on general
analysis of loop effects. Our assumption that the string field
action is given by the partition function on the
world sheet with the topology of an annulus is therefore justifiable.\\

\bf{Acknowledgements}: \rm We thank R. Rashkov for suggesting and
participating in this work in the early stages and for valuable
correspondences. Fruitful discussions with N. Hambli, Taejin Lee,
P. Matlock and H. Ooguri are gratefully acknowledged. We also
thank M. Alishahiha for useful correspondences.
\\

\end{document}